# ENZYMATIC LOGIC GATES WITH NOISE-REDUCING SIGMOID RESPONSE

Valber Pedrosa,[†$] Dmitriy Melnikov,[‡] Marcos Pita,[‡#] Jan Halámek,[‡]
Vladimir Privman,[‡] Aleksandr Simonian,[†] and Evgeny Katz[‡]

[†] *Materials Research and Education Center, Auburn University, Auburn, AL 36849, USA*

[$] *Institute of Bioscience, UNESP, Botucatu, Brazil*

[‡] *Department of Chemistry and Biomolecular Science, and Department of Physics, Clarkson University, Potsdam, NY 13699, USA*

[#] *Instituto de Catalisis y Petroleoquimica, CSIC, C/Marie Curie 2, 28040 Madrid, Spain*

**Abstract**

Biochemical computing is an emerging field of unconventional computing that attempts to process information with biomolecules and biological objects using digital logic. In this work we survey filtering in general, in biochemical computing, and summarize the experimental realization of an **AND** logic gate with sigmoid response in one of the inputs. The logic gate is realized with electrode-immobilized glucose-6-phosphate dehydrogenase enzyme that catalyzes a reaction corresponding to the Boolean **AND** functions. A kinetic model is also developed and used to evaluate the extent to which the performance of the experimentally realized logic gate is close to optimal.

## 1. Introduction

Recently there has been significant interest in chemical [1-6] and/or biochemical [7-11] information processing, including that based on enzymatic reactions [12]. Enzymatic reactions have been shown to mimic digital logic gate functions [13-15] and elementary arithmetic operations [16] as well as "networked" in binary logic circuits [17-20]. These reactions rewritten in terms Boolean logic, promise new applications such as "smart" multiple-input sensors of the "digital" (threshold) nature, particularly for biomedical applications [21-28]. For example, the output chemical concentration reaching a certain logic-**1** value, could signal that an action is needed, whereas concentrations at logic-**0** would indicate "no action."

The biochemical networks are, however, prone to noise build-up and amplification. There are many sources of noise in biochemical networks. The most ubiquitous is probably due to the random deviations in input concentrations that result in the noise in the output values [29]. The noise amplification can be quite severe in these systems and can actually prevent interconnecting of more than a couple of logic gates into networks [30]. Thus noise suppression becomes an important issue in the design of the biochemical logic elements and networks.

In general, in order to minimize or prevent noise build-up, we have to pass the signals through filters that have a sigmoid response profile with small slopes at and around logic points. However, such a response is generally not easy to achieve in biochemical enzyme-based reactions, even though sigmoid response is observed in natural systems [31-33]. This is because typical response of (bio)chemical systems is convex with the output initially linear for small inputs but reaching saturation for large inputs due to limited reaction rates and reactant supply.

Recently it has been demonstrated [34] that the addition of a reagent fast consuming a fraction of the output of the enzymatic reaction at low inputs, creates an inflection point in the response curve so that it becomes sigmoid. However, the added reactant is irreversibly consumed during reaction, which makes the whole system usable only once, unless this "filtering agent" is recovered by, for example, electrochemical means. The advantages of such an approach include: feasibility, ability to suppress the noise in a controllable manner, and since it acts on the output directly, the response can be sigmoid in both inputs, though the latter property has not been experimentally demonstrated thus far.

An alternative approach has been to use systems with a "built-in" sigmoid response due to their self-promoter response to at least one of the inputs. Many allosteric enzymes have this property, but no experimental implementations of their bulk-solution action for biochemical information processing have been reported so far. We recently initiated a study of an electrode-immobilized enzyme the function of which realizes a binary **AND** gate with a sigmoid response in one of the inputs [35]. Here we will highlight these results, which also received popular-press attention [36], as an illustration of the biochemical filtering as a noise-amplification prevention methodology.

Our **AND** logic function (gate) is based on electrode-immobilized enzyme glucose-6-phosphate dehydrogenase (G6PDH). The biocatalytic reaction

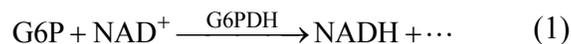
$$\text{G6P} + \text{NAD}^+ \xrightarrow{\text{G6PDH}} \text{NADH} + \cdots \qquad (1)$$

has two inputs: glucose-6 phosphate (G6P) and cofactor nicotinamide adenine dinucleotide ($\text{NAD}^+$), and one output: the reduced cofactor (NADH); the other product ($\cdots$) in aqueous solution is 6-phosphogluconate. The process in Eq. (1), carried out with electrode-immobilized enzyme experimentally displays sigmoid behavior in one of its inputs: G6P. Section 2 provides brief description of the experimental procedure. In Section 3, we present a phenomenological kinetic model suitable for the biochemical reaction studied and for the **AND** gate that it realizes. Finally, in Section 4 we present our results for the fitting of the experimental data and describe gate optimization as well as offer a summarizing discussion.

## 2. Cyclic voltammetry measurements

The biocatalytic activity of the G6PDH-modified electrode was followed by cyclic voltammetry measurements in the presence of G6P and $\text{NAD}^+$ in the solution. The reaction (1) with the immobilized enzyme resulted in the production of the reduced cofactor, NADH, which was re-oxidized electrochemically. The oxidation of NADH was observed as the anodic current peak on the voltammograms and was taken as the output signal of the enzyme-based AND logic gate. The measured cyclic voltammograms confirmed that NADH is directly oxidized at the electrode. The set of cyclic voltammograms obtained at variable concentrations of both inputs, G6P and $NAD^+$, was used to map the response surface of the enzyme-modified electrode. The concentration ranges for both chemical inputs, G6P and $NAD^+$, were selected to demonstrate the sigmoid domain of the output signal function, as presented and discussed below.

## 3. Modeling

### 3.1. *Gate response function*

To map out the response function of the logic gate, the output signal should be measured for inputs not only at the logic-0 and 1 but also for intermediate values of the reduced concentration variables, defined as

$$x = [NAD^+](t=0)/[NAD^+]_{max}, \quad (2)$$

$$y = [G6P](t=0)/[G6P]_{max}, \quad (3)$$

$$z = [NADH](t=t_{gate})/[NADH]_{max}. \quad (4)$$

Here the notation $[\ldots]_{max}$ is used to denote the maximum concentrations corresponding to the logic-1 input values at $t = 0$, and the output at the time of the voltammogram recording, $t = t_{gate}$. Thus, if we scan inputs between 0 and the maximum, and record the corresponding output, we can map out the gate response surface, $z = F(x,y)$.

As described in our recent works [29,30], networking of a gate in a larger "circuit" for biochemical logic applications requires control of the noise buildup. The *level* of the noise largely depends on the gate's environment. However, the degree of analog noise *amplification* should be kept in check to ensure stable, scalable operation of increasingly complex networks. To this end, we would like to have a response function $F(x,y)$ that has small gradients at *all* the logic points. In other words, the noise properties of the gate can be quantified by the absolute values of the gradients, $|\vec{\nabla}F|_{00,01,10,11}$, of the function $F$ at the logic points. With the largest

– 4 –

gradient less than 1, such a gate would actually offer analog noise spread *suppression*, i.e., incorporate a filtering feature in its function. As mentioned earlier, such a response is difficult to achieve in single, presently realized enzymatic gates, which typically have the largest gradient greater than 1 and amplify analog noise. Our results [35] have yielded a realization of a response surface "sigmoid" in one of the two inputs; see Fig. 1.

### 3.2. *Rate equations for the biocatalytic reaction*

The biocatalytic process in (1) is not a direct reaction. It involves several steps and possible competing pathways, the precise kinetics of which is not fully understood, especially for the electrode-immobilized enzyme as the biocatalyst. We use a phenomenological rate equation approach [35] within a simplified, few-parameter description, which bypasses the kinetic issues of the experimentally observed "self-promoter" property of G6P as a substrate. The rate equation for G6P, denoted by $[G6P](t) = G(t)$, is

$$dG/dt = -[\alpha + \beta(G_0 - G)]GM . \qquad (5)$$

We assume that the dominant reaction pathway is the one corresponding to this substrate being captured by the enzyme, of concentration $[G6PDH](t) = M(t)$. The reaction rate is initially proportional to $\alpha G(t)M(t)$, while at later times the rate increases proportionately to the amount of the consumed substrate (the "self-promoting" effect), $\alpha \to \alpha + \beta[G_0 - G(t)]$, where $G_0 = G(0)$ is the initial concentration of the input G6P.

The resulting complex of concentration $C(t)$, then combines with $NAD^+$, of concentration $[NAD^+](t) = N(t)$ to yield the product $[NADH](t) = P(t)$, as well as to restore the biocatalyst. We will use this rather simplified description to write the remaining rate equations for our system:

$$dC/dt = [\alpha + \beta(G_0 - G)]GM - \gamma NC , \qquad (6)$$

$$dM/dt = -[\alpha + \beta(G_0 - G)]GM + \gamma NC , \qquad (7)$$

$$dP/dt = -dN/dt = \gamma NC . \qquad (8)$$

Here $\alpha$, $\beta$ and $\gamma$ are adjustable parameters (rates) which are varied to fit experimental data (Section 4) and to explore the noise-amplification properties of our biocatalytic system.

## 4. Data analysis and gate-function properties

The experimental response surface is shown in Fig. 1; the fitting of these data with the rate equations (5)-(8) is also shown. Since the experimental response function is rather noisy, the fitting was performed as a weighted non-linear least-squares fit of the data with emphasis in the region of small [G6P], for which the self-promoter property is observed. This yielded $\alpha = 0.03\,\text{mM}^{-1}\text{s}^{-1}$, $\beta = 42\,\text{mM}^{-2}\text{s}^{-1}$, $\gamma = 1.05\,\text{mM}^{-1}\text{s}^{-1}$. Note that an effective initial enzyme concentration, $M_0$, can also be regarded as another variational parameter. This is because the applied enzyme is only partially immobilized (the rest is washed out after the immobilization step), giving rise to an unknown load of the enzyme at the electrode interface, which is usual for this kind of immobilization procedures [37]. Due to this, in the fitting procedure we fixed the effective enzyme concentration at $M_0 = 1\,\mu\text{M}$. We also found that the noise optimization results discussed below weakly depend on the value of $M_0$ provided that it changes within a factor of 2.

We then computed [35] the gradients at the four logic points as functions of the effective enzyme concentration and reaction time (Figure 2). The maximum of these gradients in the parameter range covered is at the logic-**10** point which is also confirmed by visual inspection of the response surfaces in Figure 1: One can see that the largest variations in the output are indeed around the logic-**10**. Slopes at other logic points are predictably smaller; in particular, slope at the logic-**01** is close to zero due to the sigmoid response. Interestingly, our randomly selected (for experimental convenience) values of the parameters yield the value of the maximum slope (noise amplification factor) ~1.16, which is already very good [35].

By changing the enzyme activity and/or the gate time, while keeping all the other parameters fixed, we can further "optimize" the gate performance. Specifically, we need to minimize the noise amplification factor, which in our case is defined by the gradient at the logic-**10**. As seen in Figure 2, there is a broad region of reaction times longer than ~ 200 s and enzyme activities comparable or larger than in our system, which yields practically no noise amplification, i.e., the largest gradient is very close to 1, with values as low as ~ 1.05 within experimental reach.

To conclude, we reviewed the experimental realization and numerical performance analysis of an enzymatic AND gate with a noise-reducing "sigmoid" response in one of the inputs. The logic gate was realized with electrode-immobilized glucose-6-phosphate dehydrogenase enzyme. The studied enzymatic reaction was found to have a relatively small degree of analog noise amplification for the selected experimental parameters, which turned out to be close to the theoretically identified optimal regime. Let us compare the reviewed systems to the recently developed logic filter [34] for which the sigmoid profile is achieved by an addition of a reactant interacting with the output signal. The generality (versatility) of the built-in mechanism is unclear until more experiments are carried out with enzymes that demonstrate nonlinear response to substrates or other regulator molecules, e.g., [38]. Therefore the challenge of developing enzymatic systems with built-in sigmoid response might be generally addressable only with the help of added network elements involving additional chemicals, rather than at the level of individual enzyme gates.

**Acknowledgements**

The Clarkson team acknowledges support by NSF (CCF-0726698 and CCF-1015983, DMR-0706209), by ONR (N00014-08-1-1202), and by SRC (2008-RJ-1839G). The Auburn team acknowledges support from the USDA-CSREES (2006-34394-16953). The work in Spain (M.P.) was funded by Ramon y Cajal program, MICINN. Additionally, this material is based on the work (A.S.) which was supported by NSF, while working (A.S.) at the Foundation. Any opinion, finding, and conclusions or recommendations expressed in this material are those of the authors and do not necessarily reflect the views of NSF, DoD, or US Government.

# References

[1] De Silva, A.P., Uchiyama, S., Vance, T.P., Wannalerse, B. (2007). A supramolecular chemistry basis for molecular logic and computation. *Coord. Chem. Rev.* **251**, 1623-1632.

[2] Credi, A. (2007). Molecules that make decisions. *Angew. Chem. Int. Ed.* **46**, 5472-5475.

[3] Pischel, U. (2007). Chemical approaches to molecular logic elements for addition and subtraction. *Angew. Chem. Int. Ed.* **46**, 4026-4040.

[4] Szacilowski, K. (2008). Digital information processing in molecular systems. *Chem. Rev.* **108**, 3481-3548.

[5] Pischel, U. (2010). Digital operations with molecules - Advances, challenges, and perspectives. *Austral. J. Chem.* **63**, 148-164.

[6] Andreasson, J., Pischel, U. (2010). Smart molecules at work-mimicking advanced logic operations. *Chem. Soc. Rev.* **39**, 174-188.

[7] Benenson, Y., Gil, B., Ben-Dor, U., Adar, R., Shapiro, E. (2004). An autonomous molecular computer for logical control of gene expression. *Nature* **429**, 423-429.

[8] Shapiro, E., Gil, B. (2007). Biotechnology - Logic goes in vitro. *Nature Nanotechnol.* **2**, 84-85.

[9] Benenson, Y. (2009). Biocomputers: from test tubes to live cells. *Molecular Biosystems* **5**, 675-685.

[10] Stojanovic, M.N., Stefanovic, D., LaBean, T., Yan, H. (2005). Computing with nucleic acids. In: Willner, I., Katz, E. (Eds.), *Bioelectronics: From Theory to Applications*, Wiley-VCH, Weinheim, pp. 427-455.

[11] Win, M.N., Smolke, C.D. (2008). Higher-order cellular information processing with synthetic RNA devices. *Science* **322**, 456-460.

[12] Katz, E., Privman, V. (2010). Enzyme-based logic systems for information processing. *Chem. Soc. Rev.* **39**, 1835-1857.

[13] Baron, R., Lioubashevski, O., Katz, E., Niazov, T., Willner, I. (2006). Logic gates and elementary computing by enzymes. *J. Phys. Chem. A* **110**, 8548-8553.

[14] Strack, G., Pita, M., Ornatska, M., Katz, E. (2008). Boolean logic gates using enzymes as input signals. *ChemBioChem* **9**, 1260-1266.

[15] Zhou, J., Arugula, M.A., Halámek, J., Pita, M., Katz, E. (2009). Enzyme-based universal NAND and NOR logic gates with modular design. *J. Phys. Chem. B* **113**, 16065-16070.

[16] Baron, R., Lioubashevski, O., Katz, E., Niazov, T., Willner, I. (2006). Elementary arithmetic operations by enzymes: A model for metabolic pathway based computing. *Angew. Chem. Int. Ed.* **45**, 1572-1576.

[17] Niazov, T., Baron, R., Katz, E., Lioubashevski, O., Willner, I. (2006). Concatenated logic gates using four coupled biocatalysts operating in series. *Proc. Natl. Acad. Sci. USA* **103**, 17160-17163.

[18] Strack, G., Ornatska, M., Pita, M., Katz, E. (2008). Biocomputing security system: Concatenated enzyme-based logic gates operating as a biomolecular keypad lock. *J. Am. Chem. Soc.* **130**, 4234-4235.

[19] Privman, M., Tam, T.K., Pita, M., Katz, E. (2009). Switchable electrode controlled by enzyme logic network system: Approaching physiologically regulated bioelectronics, *J. Am. Chem. Soc.* **131**, 1314-1321.

[20] Tam, T.K., Pita, M., Katz, E. (2009). Enzyme logic network analyzing combinations of biochemical inputs and producing fluorescent output signals: Towards multi-signal digital biosensors. *Sens. Actuat. B* **140**, 1-4.

[21] Margulies, D., Hamilton, A.D. (2009). Digital analysis of protein properties by an ensemble of DNA quadruplexes. *J. Am. Chem. Soc.* **131**, 9142-9143.

[22] Adar, R., Benenson, Y., Linshiz, G., Rosner, A., Tishby, N., Shapiro, E. (2004). Stochastic computing with biomolecular automata. *Proc. Natl. Acad. USA* **101**, 9960-9965.

[23] May, E.E., Dolan, P.L., Crozier, P.S., Brozik, S., Manginell, M. (2008). Towards de novo design of deoxyribozyme biosensors for GMO detection. *IEEE Sens. J.* **8**, 1011-1019.

[24] Manesh, K.M., Halámek, J., Pita, M., Zhou, J., Tam, T.K., Santhosh, P., Chuang, M.-C., Windmiller, J.R., Abidin, D., Katz, E., Wang, J. (2009). Enzyme logic gates for the digital analysis of physiological level upon injury. *Biosens. Bioelectron.* **24**, 3569-3574.

[25] Pita, M., Zhou, J., Manesh, K.M., Halámek, J., Katz, E., Wang, J. (2009). Enzyme logic gates for assessing physiological conditions during an injury: Towards digital sensors and actuators. *Sens. Actuat. B* **139**, 631-636.

[26] Halámek, J., Windmiller, J.R., Zhou, J., Chuang, M.-C., Santhosh, P., Strack, G.,

Arugula, M.A., Chinnapareddy, S., Bocharova, V., Wang, J., Katz, E. (2010). Multiplexing of injury codes for the parallel operation of enzyme logic gates. *Analyst* **135**, 2249-2259.

[27] Wang, J., Katz, E. (2010). Digital biosensors with built-in logic for biomedical applications – Biosensors based on biocomputing concept. *Anal. Bioanal. Chem.* **398**, 1591-1603.

[28] Windmiller, J.R., Strack, G., Chuan, M.-C., Halámek, J., Santhosh, P., Bocharova, V., Zhou, J., Katz, E., Wang, J. (2010). Boolean-format biocatalytic processing of enzyme biomarkers for the diagnosis of soft tissue injury. *Sens. Actuat. B* **150**, 285-290.

[29] Privman, V., Strack, G., Solenov, D., Pita, M., Katz, E. (2008). Optimization of enzymatic biochemical logic for noise reduction and scalability: How many biocomputing gates can be interconnected in a circuit? *J. Phys. Chem. B* **112**, 11777-11784.

[30] Privman, V., Arugula, M.A., Halámek, J., Pita, M., Katz, E. (2009). Network analysis of biochemical logic for noise reduction and stability: A system of three coupled enzymatic AND gates. *J. Phys. Chem. B* **113**, 5301-5310.

[31] Buchler, N.E., Gerland, U., Hwa, T. (2005). Nonlinear protein degradation and the function of genetic circuits. *Proc. Natl. Acad. Sci. USA* **102**, 9559-9564.

[32] Setty, Y., Mayo, A.E., Surette, M.G., Alon, U. (2003). Detailed map of a cis-regulatory input function. *Proc. Natl. Acad. Sci. USA* **100,** 7702-7707.

[33] Alon, U. (2007). *An Introduction to Systems Biology. Design Principles of Biological Circuits*, Boca Raton, Florida: Chapman & Hall/CRC Press.

[34] Privman, V., Halámek, J., Arugula, M. A., Melnikov, D., Bocharova, V., Katz, E. (2010). Biochemical filter with sigmoidal response: Increasing the complexity of biomolecular logic. *J. Phys. Chem. B* in press.

[35] Privman, V., Pedrosa, V., Melnikov, D., Pita, M., Simonian, A., Katz, E. (2009). Enzymatic AND-gate based on electrode-immobilized glucose-6-phosphate dehydrogenase: Towards digital biosensors and biochemical logic systems with low noise. *Biosens. Bioelect.* **25**, 695-701.

[36] Researchers from Clarkson University report details of new studies and findings in the area of biosensors and bioelectronics. *Electronics Newsweekly* (Atlanta, GA), Issue: Jan.

13, 2010, Page: 161 (online version at http://www.verticalnews.com/article.php?articleID=3010969).

[37] Willner, I., Katz, E. (2000). Integration of layered redox-proteins and conductive supports for bioelectronic applications. *Angew. Chem. Int. Ed.* **39**, 1180-1218.

[38] Rabinowitz, J. D., Hsiao, J. J., Gryncel, K. R., Kantrowitz, E. R., Feng, X.-J. (2008). Dissecting Enzyme Regulation by Multiple Allosteric Effectors: Nucleotide Regulation of Aspartate Transcarbamoylase. *Biochemistry* **47**, 5881-5888.

## Nomenclature

| | | |
|---|---|---|
| G6P | | Glucose-6 phosphate |
| G6PDH | | Glucose-6-phosphate dehydrogenase from *Leuconostoc mesenteroides* (E.C. 1.1.1.49) |
| $NAD^+$ | | β-Nicotinamide adenine dinucleotide cofactor |
| NADH | | β-Nicotinamide adenine dinucleotide reduced cofactor |
| $x$ | | Logic input |
| $y$ | | Logic input |
| $z$ | | Logic output |
| $F(x,y)$ $(= z)$ | | Gate response surface |
| $|\nabla F|$ | | Gradient of the response surface |
| $G(t)$ | | Concentration of G6P |
| $M(t)$ | | Concentration of G6PDH |
| $N(t)$ | | Concentration of $NAD^+$ |
| $P(t)$ | | Concentration of NADH |
| $C(t)$ | | Concentration of the intermediate complex |
| $\alpha$ | 0.03 $mM^{-1}s^{-1}$ | Reaction rate |
| $\beta$ | 42 $mM^{-2}s^{-1}$ | Reaction rate |
| $\gamma$ | 1.05 $mM^{-1}s^{-1}$ | Reaction rate |
| $M_0$ | 1 µM | Effective initial concentration of the enzyme |

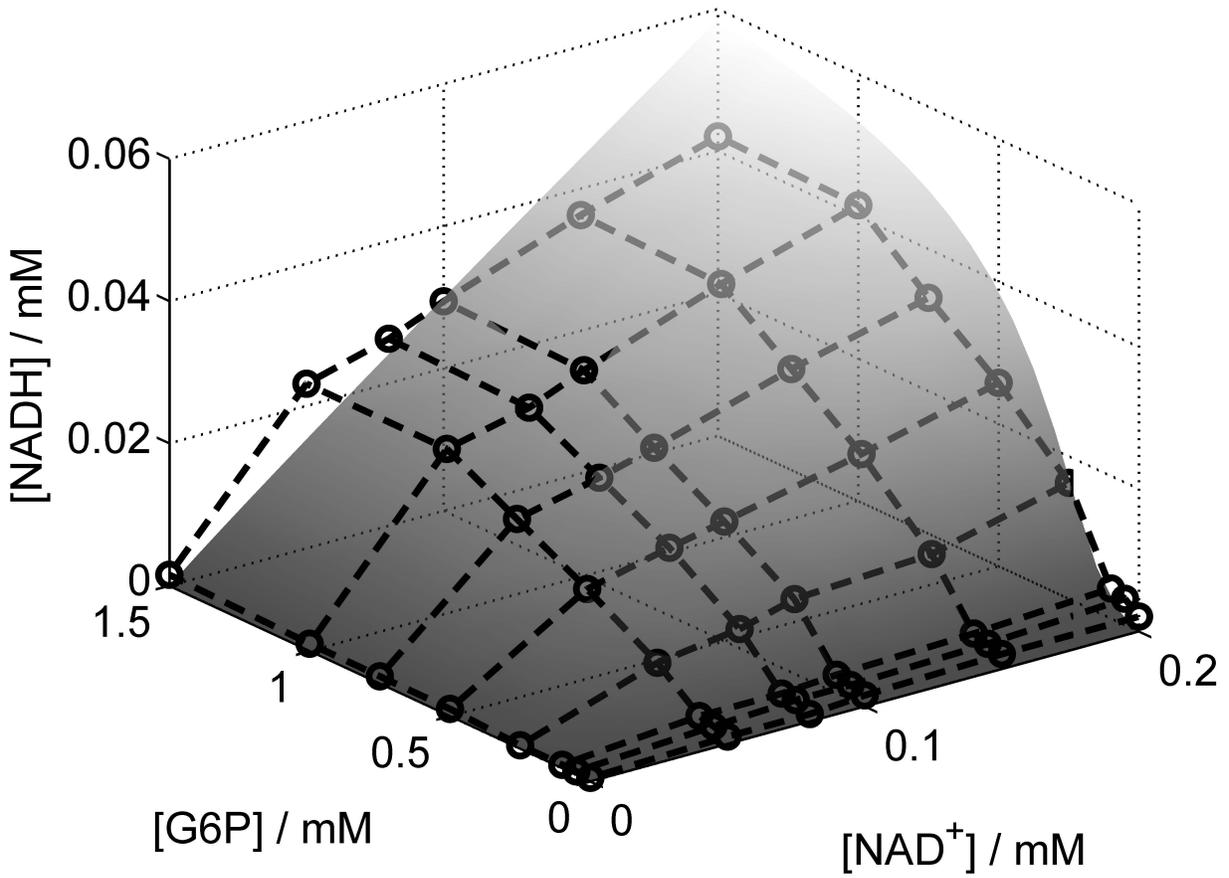

**Figure 1:** Response surface at $t_{gate}$ = 300 s. Open circles are the experimental data while continuous surface is the computed data fit. Dashed lines between circles are for guiding the eye. Note that in the [NAD$^+$] direction, the fit does not give a good description of the measured data as our phenomenological kinetic model (5)-(8) focuses on description of the response to the G6P concentration.

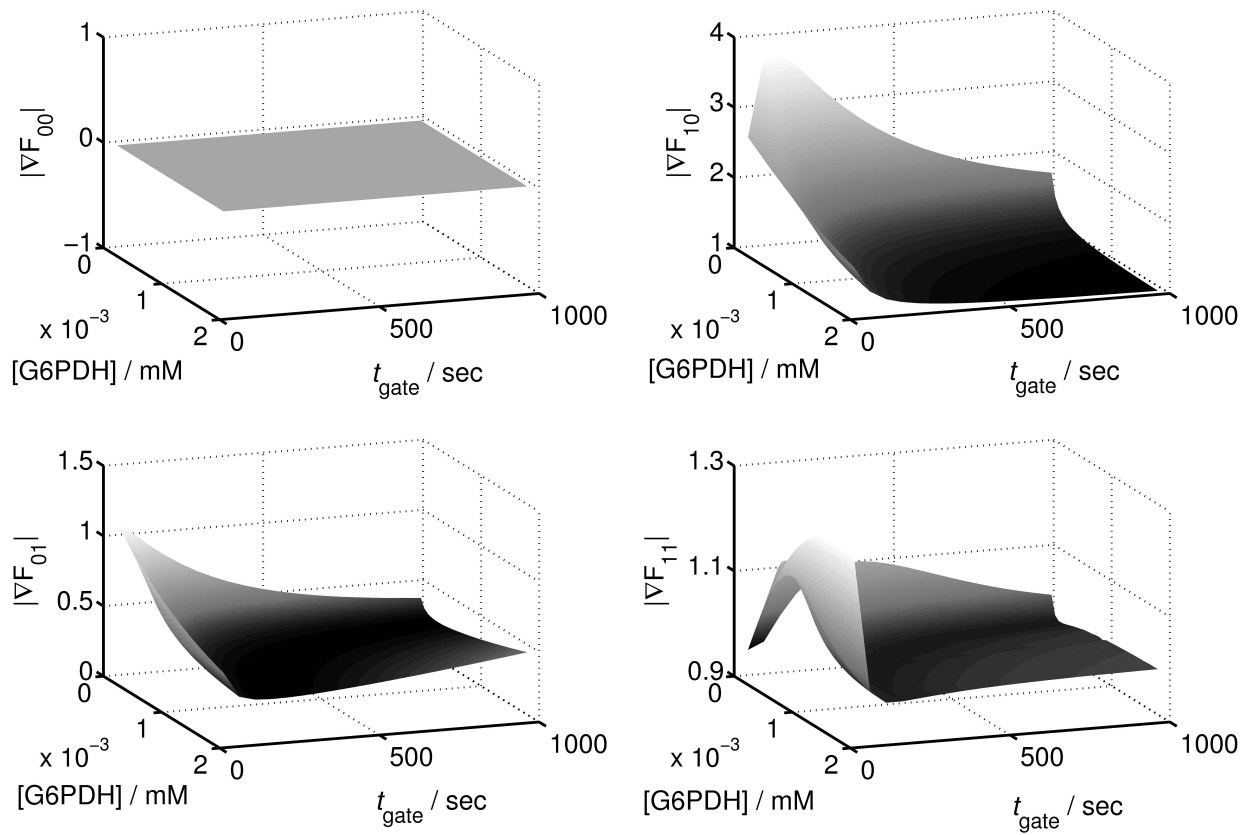

**Figure 2:** Gradients of the response surface at four logic points vs. the reaction time and [G6PDH]. At our experimental conditions ($t_{\text{gate}} = 300$ sec, $M_0 = 1$ μM) the maximum gradient is ~ 1.16.